\input{aipcheck}

\documentclass[
    ,final            
  ]
  {aipproc}

\layoutstyle{6x9}
\usepackage{natbib}
\usepackage{astro_bib_macro}

\begin{document}

\title{The inner environment of protoplanetary disks with
   near infrared spectro-interferometry}


\author{E. Tatulli}{
  address={}
}

\begin{abstract}
In this paper, I review how optical spectro-interferometry
has become a particularly well suited technique to study the close
environment of young stars, by spatially resolving both their IR continuum
and line emission regions. I summarize in which ways optical
interferometers have brought major insights about our understanding of
the inner part of circumstellar disks, a region in which the
first stages of planet formation are thought to occur. In particular, I
emphasize how new methods are now enabling to probe the hot gas
emission, in addition to the circumstellar dust. 
\end{abstract}

\maketitle


\section{Introduction} \label{sec_intro}
\begin{figure}
\includegraphics[width=\textwidth]{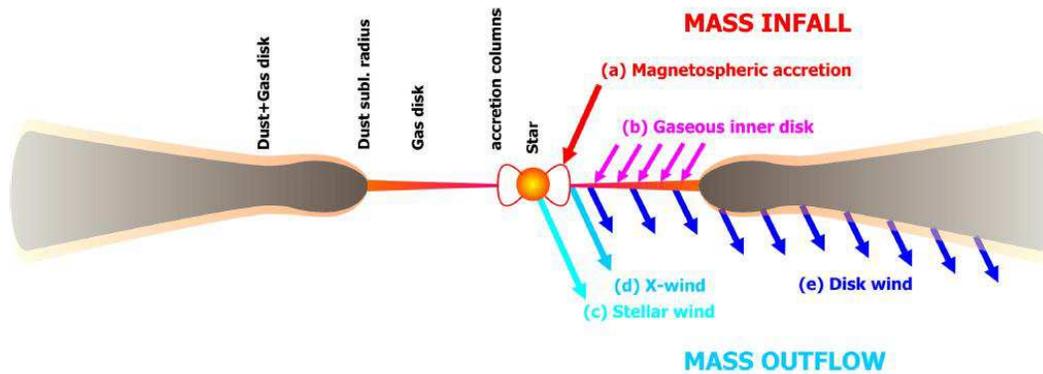}
\caption{\footnotesize{Sketch of the inner environment around young stellar
  objects, from \citep{kraus_1}. See text for detailed description.}}
\label{fig1}
\end{figure}

Observing the protoplanetary disks around young stars is a key issue
to understand the first steps of planet formation mechanisms. Such
processes are occurring in the very inner environment of the central
star, at distances of a few Astronomical Units. The representation
that we have today of this environment is sketched in
Fig. \ref{fig1}, and is basically composed of i) magnetically-driven
columns of gas accreting on the central star, ii) a gaseous dust-free
rotating disk, iii) a dusty disk which inner rim is located at the
dust sublimation radius; and iv) potentially outflowing winds. Observational clues that we can obtain of the inner part of the
protoplanetary disks are twofold: \\
{\it From their continuum infrared excess}, that arises
from the emission of the hot circumstellar dust and gas. It will give information about the
structure/geometry of the disk as well as about its composition (e.g. grain growth, radial/vertical distribution, mineralogy);\\
{\it From their infrared emission lines}, in particular the
hydrogen lines, that can originate from mainly two different
mechanisms: whether magnetospheric accretion along the accreting columns of
gas \citep{hartmann_1} or through magnetically-driven outflows \citep{shu_1,
  casse_1, sauty_1}. \\ 
In order to characterize these mechanisms unambiguously, one needs
both spatial and spectral resolution to localize and separate the
continuum and line emission regions. At distances of the first
stellar formation regions ($\sim 150$pc), 1AU corresponds to a
angular distance of $\sim 6$mas, a resolution that only
interferometric techniques can achieve. Furthermore, at such
distances from the star, the temperature at the inner region of young
stars is roughly between a few 100K and a few 1000K, that is radiating
at near infrared wavelengths. As a result, near infrared
spectro-interferometry  which provides both the spatial and spectral
resolution required at the desired wavelengths
appears to be a technique perfectly suited to
observe the inner environment of protoplanetary disks.
\section{Origin of the near infrared excess: the K-band size-luminosity relationship}\label{sec_dust}
By observing young stars at near infrared wavelengths, interferometry
has enabled to {\it locate} the emission region responsible for the
continuum infrared excess, and potentially constraint its
structure. And such dimensional constraints appeared to be critical to
unveil the physical origin of this emission, both for low (T Tauri)
and intermediate (Herbig Ae/Be) mass young stars. 
\subsection{Herbig Ae/Be stars: thermal emission of the dusty inner rim}
\begin{figure}
\includegraphics[width=\textwidth]{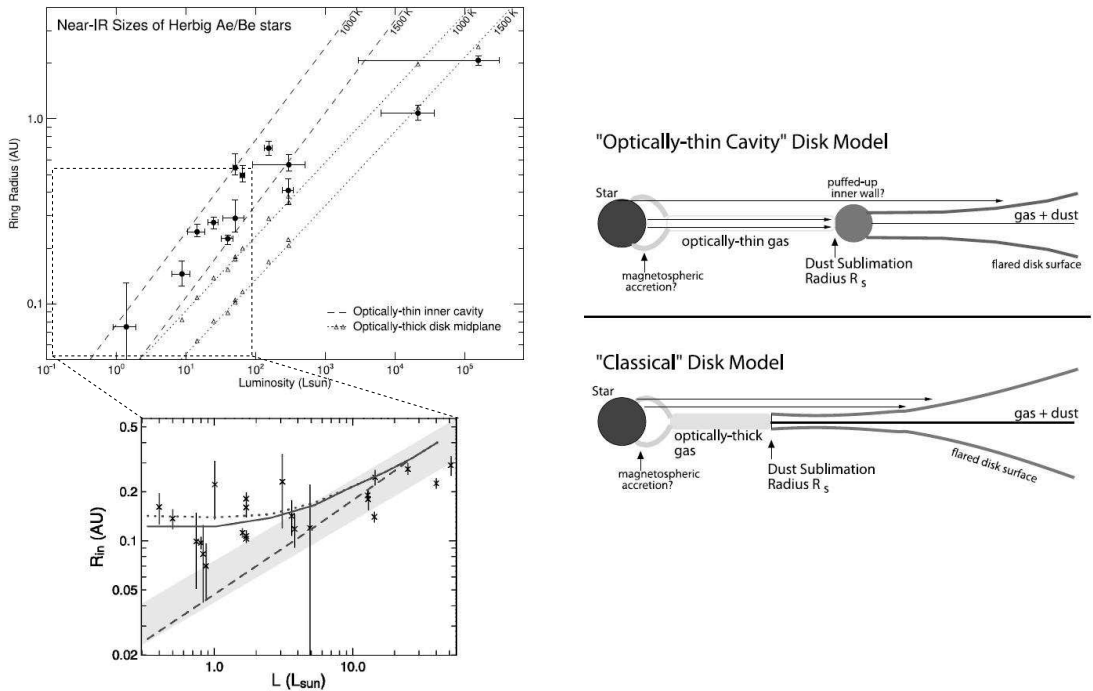}
\caption{\footnotesize{Left: K-band interferometric size-luminosity
    relationship for intermediate \citep{monnier_1} and low
    \citep{pinte_1} mass young stars. We can see in the bottom plot
    that for the T Tauri regime, considering the thermal emission only
    (dashed line) does not reproduce the correlation whereas taking
    into account both the thermal and scattered light emission with
    the same disk model (solid
    line) does. Right: schematic representation of the inner
    environment of young stars, with respectively optically thin (top) and
    optically thick (bottom) material within the dust sublimation
    radius (from \citep{monnier_1}).}}
\label{fig_sizelum}
\end{figure}

\citet{monnier_1, vinkovic_1} have shown on a sample of  Herbig
Ae/Be stars that the interferometric size of the K-band emission was
correlated with the star luminosity, as illustrated on Fig. \ref{fig_sizelum} (left). From
this correlation they  have demonstrated that the
near infrared excess of such stars was -- at the exception of the
most luminous ones -- arising  from the thermal emission of the inner
part of the dusty circumstellar disk, located at the dust sublimation
radius, assuming  that the dust is in equilibrium with the radiation
field (ie $R_{sub} \sim 0.5
\sqrt{\frac{Q_{abs}(T_{\ast})}{Q_{abs}(T_{sub})}}\left(\frac{T_{\ast}}{T_{sub}}\right)^2R_{\ast}$,where
$T_{sub}$ is the dust sublimation temperature, roughly $T \sim
1500$K for silicates). If this scenario works well for Herbig Ae stars
and late Be, it however fails to interpret the size of the infrared
excess emission region for the early Be, the inner rim being too close
to the star regarding their high luminosity. 
In this case, one likely interpretation is that
the gas inside the dust sublimation radius is optically thick to the
stellar radiation, hence shielding a fraction of the stellar light and
allowing the dusty inner rim to move closer to the star, as sketched
on Fig. \ref{fig_sizelum} (right).
\subsection{T Tauri stars:  a strong contribution of the scattered light}
Together with the last improvements of interferometers in terms of
sensitivity, it is only recently that the same kind of study could
have been performed on the less luminous T Tauri stars. And the
results that have been obtained were somewhat surprising, the size of
the NIR emission being {\it larger} than predicted \citep{akeson_1,
  eisner_1, eisner_2}. Many  hypothesis were invoked such as lower
sublimation  temperature $T_{sub} \sim 1000$K, fast dissipation of the
inner disk, magnetospheric radii bigger than dust sublimation ones
hence defining the location of the inner rim... until  \citet{pinte_1}
have shown that as long as the luminosity of the star decreases, the
contribution of the scattered light, in addition to that of the thermal
emission, could not neglected anymore.  As a consequence, as shown inf
Fig. \ref{fig_sizelum}, these authors
have convincingly demonstrated that the model of the inner disk located at the dust sublimation radius was
holding for the T Tauri regime as well, and that no alternative
scenario was required as long as the radiative transfer in the disk
was thoroughly studied (thermal + scattered light).

\section{Resolving the hot gas continuum emission}
Though dust is mostly dominating the near infrared continuum emission
of young stars, there are some cases, especially for stars where the accretion
rate is  high enough -- roughly $> 10^{-7}M_{\odot}$/yr -- where the
contribution of the dust-free hot gaseous component to the NIR excess is not
negligible. Since the dust-free gas is located between the star and
the dust sublimation radius, we expect the region of emission to be
{\it hotter} and {\it more compact} than that of the dusty inner
rim. As a consequence, going towards shorter wavelengths or longer
baselines appear to be well suited strategies to probe this region.\\
\begin{figure}
\includegraphics[width=0.8\textwidth]{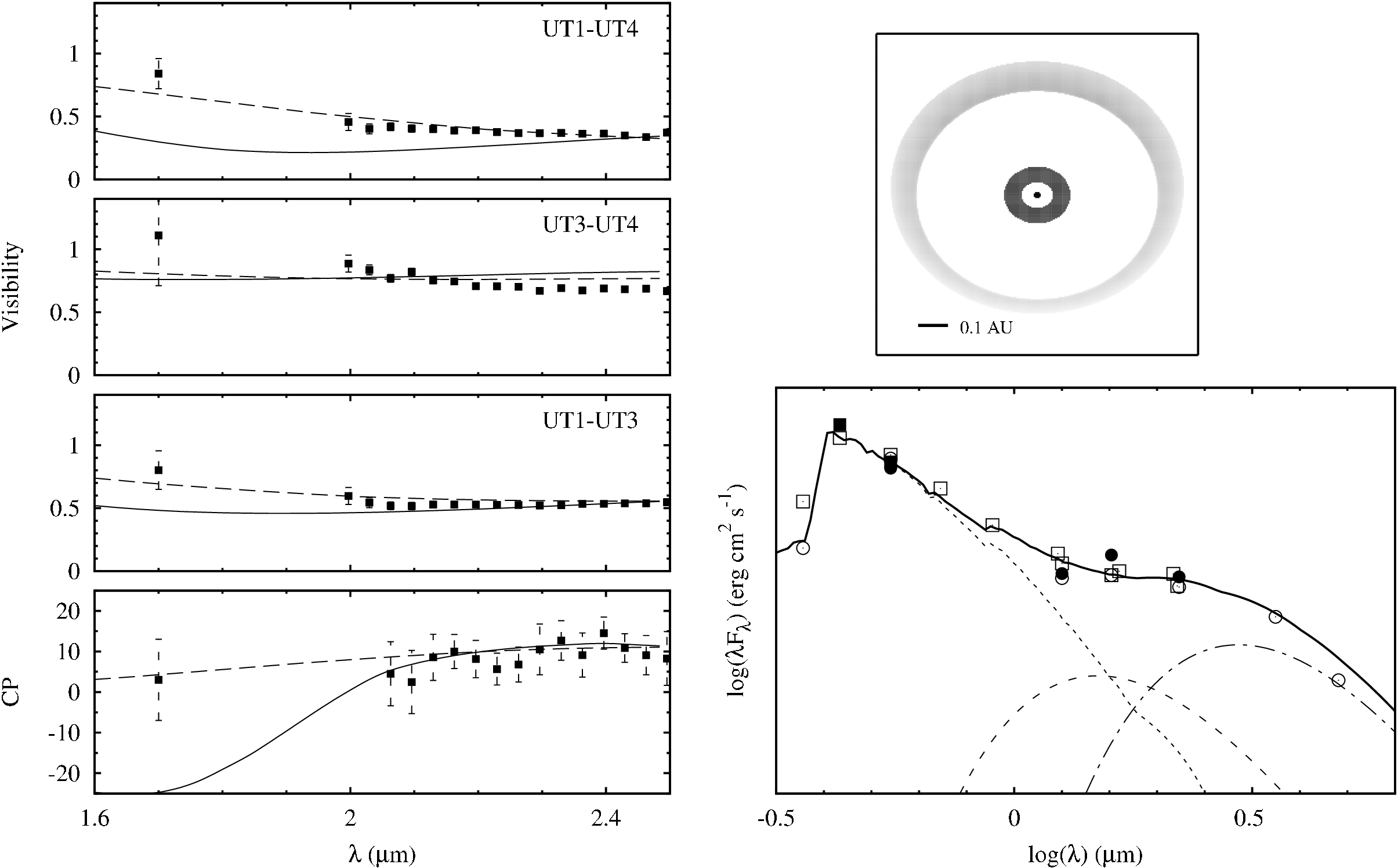}
\caption{\footnotesize{Left: H and K band visibilities and closure
    phases of MWC758, from \citet{isella_1}. The solid line represents
    the model of dust inner rim only, the dashed one being the
    unresolved point + dust model as shown in the right top
    corner. Below is shown the SED, well reproduced by the star
    (dotted line) + unresolved point (dashed line) + dust (dash-dotted
    line) model.}}
\label{fig_gascont}
\end{figure}
In this framework, \citet{isella_1} have observed the Herbig Ae star
MWC 758 with the AMBER
instrument on the VLTI, both in the H and K bands. They have shown that, if the K band observations alone are well interpreted by the classical
dusty puffed-up inner rim ($T_{sub} =1400$K, $R_{in}=0.34$AU), it
fails to reproduce the H band observations for which the emission is
less resolved than expected by this model. Furthermore, with this single
model, the SED can be not fitted successfully, showing a lack of
energy in the H band. Conversely, by adding to the model the presence
of an unresolved hotter component (of $T_{g} = 2500$K), they managed
to reproduce both the H and K bands measurements jointly. Note that
this changes slightly the parameters of the dusty rim ($T_{sub}
=1300$K, $R_{in}=0.40$AU). What was then the physical interpretation
for this unresolved component? Given the temperature and the size
($\le 0.1$AU) of the emission region, it is likely that AMBER has
directly probed the hot gas accreting  close to to the
star. And indeed, models of accreting gas developed by
\citet{muzerolle_1} (assuming an accretion rate of $\sim
2.10^{-7}M_{\odot}/yr$  from the star's $\mathrm{Br}\gamma$
luminosity), allow as well to satisfactory fit the shape of the SED by
filling the lack of energy in the H band (see  Fig. \ref{fig_gascont}), hence
reinforcing this interpretation.\\
Somewhat similar strategy was used by \citet{eisner_3} who observed
different Herbig Ae/Be stars with the KecK interferometer, using moderate
spectral dispersion (R=25) within the K-band. They hence have found that for
several stars of their sample, single-temperature ring could
not reproduce the data well, and that models incorporating radial
temperature gradients or two rings should be preferred, supporting the
view that the near-IR emission of Herbig Ae/Be sources can arise from
both hot circumstellar dust and gas. For example, the interferometric
data of AB Aur require the presence of  one dust inner rim together
with a  more compact and hotter component ($T \sim 2000$K) interpreted
as coming from the hot dust-free inner gas. And as a matter of fact,
this scenario was confirm by \citet{tannirkulam_1} who observed the
same star with the very long baselines ($300$m) of the CHARA
interferometer, hence probing smaller emission region and showing the
need of adding smooth hot ($T > 1900$K) emission inside the dust inner
rim, contributing to $65\%$ of the K-band excess, as summarized in
Fig. \ref{fig_gascontabaur}.
\begin{figure}
\includegraphics[width=0.9\textwidth]{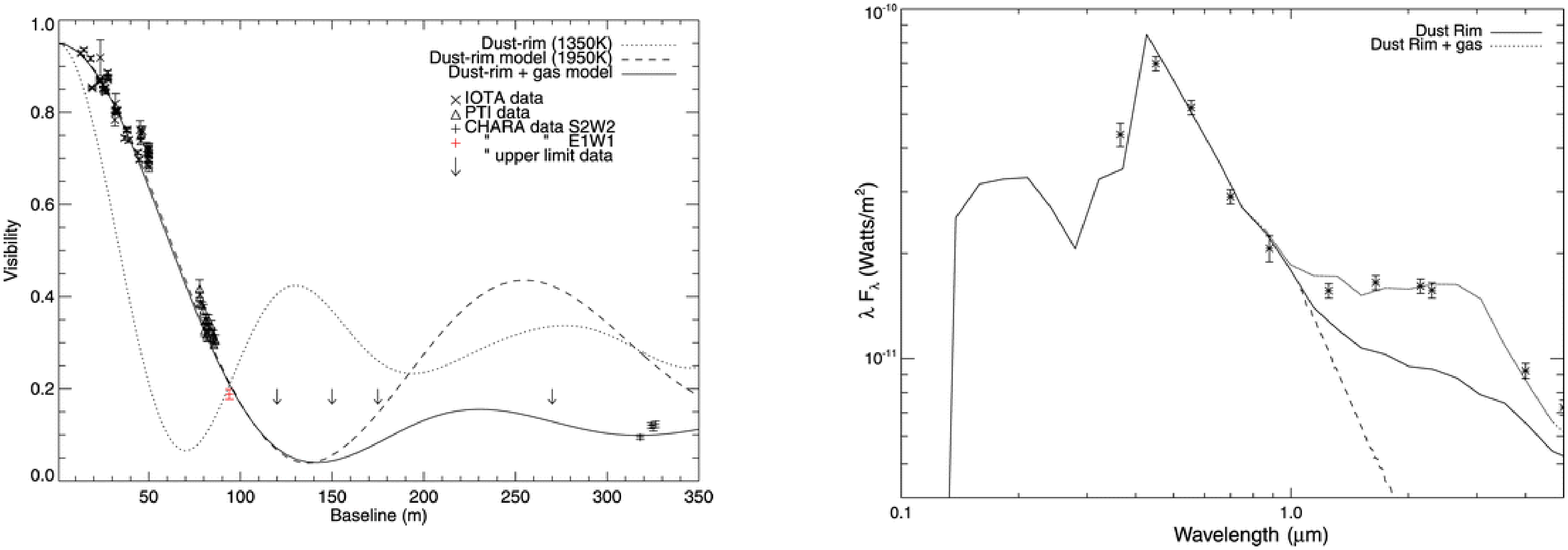}
\caption{\footnotesize{Left: Visibility of AB Aur as a function of the
    projected baseline (from \citet{tannirkulam_1}). If the dust rim
    model alone succeeds in fitting the shorter baselines up to 100m,
    the interferometric data at 300m require the presence of a compacter
    component inside the sublimation radius, interpreted as arising
    from the hot dust-free inner gas, this extra component being
    responsible for  $65\%$ of the K-band excess, as shown in the SED (right).}}
\label{fig_gascontabaur}
\end{figure}

\section{The origin of NIR emission lines}
One major achievement in interferometry in the past years is the
capacity of spectrally dispersed the interferogram with resolution
high enough (AMBER/VLTI: R=1500, and very recently KecKI: R=1700, that
is some $100$km/s) to spatially resolve the lines emission regions
together with that of the continuum, that is to directly probe the
gas which constitutes $99\%$ of the mass of the circumstellar matter. 

\subsection{The origin of
  $\mathrm{Br}\gamma$ emission in Herbig Ae/Be stars: probing the
  accretion/ejection phenomena: }
\begin{figure}
\includegraphics[width=0.95\textwidth]{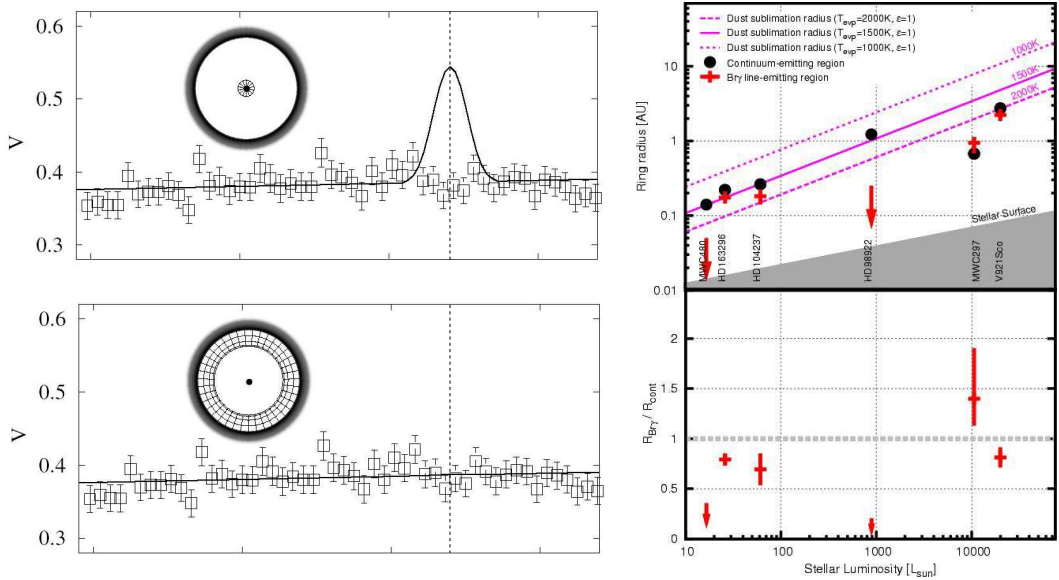}
\caption{\footnotesize{Left: visibility around the
    $\mathrm{Br}\gamma$  line at $2.16\mu$m (indicated with a dashed line) of HD104237 and superimposed
    magnetospheric-accretion (up) and wind (down) models, from
    Tatulli et al. (2007). Size of the $\mathrm{Br}\gamma$ region for 6
    Herbig Ae/Be stars as a function of the star's luminosity, from Kraus et al. (2008).}}
\label{fig_brg}
\end{figure}

Among all the NIR emission lines that are seen in young stars, the
atomic transition of the hydrogen $\mathrm{Br}\gamma$ is the most
observed in spectro-interferometry for it is the brightest and can
therefore be studied with rather good signal to noise ratio. However,
since (i) the number of measurements remains poor in classical
interferometric observations (two or 3 baselines simultaneously) and
(ii) the spectral resolution is not high enough to {\it spectrally}
resolve the line, the interferometric measurements must be, so far,
interpreted in terms of simple geometrical models. The analysis is
done as follows: measure the size of the emitting region for both the
emission line and the surrounding continuum, then compare their
relative size to put some strong constraints on the physical
mechanisms at the origin of the emission line. Typically, as described
in introduction, two main scenarios are in balance: \\
- {\it magnetospheric accretion:} If the line is emitted in
  accreting columns of gas, then the region of emission lies roughly between
  the star and the corotation radius, that is {\it the line emission region
  is much more compact than that of the continuum} which comes from
  the dust sublimation radius, as described in previous Section. \\
- {\it outflowing winds/jets:} At the contrary, for such scenario,
  we expect the  {\it line emitting region to be of the same size or bigger
  than that of the continumm.}\\
Using the AMBER instrument, \citet{malbet_1, tatulli_1, kraus_1} have observed a
sample of Herbig Ae/Be stars that displayed strong $\mathrm{Br}\gamma$
lines and have performed for each star the geometrical analysis
described above, that is by comparing the size of the line emitting
region with respect to that of the continuum, as illustrated in Fig. \ref{fig_brg}.
As a result, if for two stars
(HD98922, MWC480) the interferometric measurements were compatible with
the magnetospheric-accretion for the origin of their
$\mathrm{Br}\gamma$ emission, the wind scenario was favored for four
of them (MWC275, MWC297, V921Sco, HD104237). Taken statistically,
these results are quite  interesting to analyze:  
at the contrary of T Tauri stars for which the direct
correlation between accretion and $\mathrm{Br}\gamma$ emission seems well
established,  in Herbig Ae/Be stars we are mostly probing outflows
phenomena from $\mathrm{Br}\gamma$ emission, this latter being
probably in this case an indirect tracer of accretion through
accretion-driven mass loss. \\

\subsection{The origin of CO overtone emission: probing the hot
  molecular gas}
Whereas $\mathrm{Br}\gamma$ is a good tracer of whether magnetospheric
accretion or outflowing winds, some other lines such as the CO
overtone emission at $2.3\mu$m are also of great interest to directly
probe the hot rotating gas. One problem however is that only a few
young stars display strong enough CO lines to be observed
with spectro-interferometry. 51 Oph is one these stars \citep{thi_1,
  berthoud_1}, and \citet{tatulli_2} have recently presented the first
interferometric observations of this young star around the CO overtone
emission, using the AMBER instrument with the
resolution of 1500. They have shown that: 
(i) the hot CO emission was resolved, located at a
distance of $0.15$AU from the star, thus in agreement with
the scenario in which the CO is emitted from the first AU of a
rotating gaseous disk \citep{thi_1}, (ii) the two first bandheads are
arising from the same emitting region, and (iii)  the adjacent continuum is located at a distance of $0.25$AU, that is
too close to the star compared to the location of the sublimation
radius, suggesting that the stellar light is shielded by the
optically thick gas hence moving the sublimation radius closer to the
star, and/or that the hot gas inside the dust sublimation radius
significantly contributes to the observed 2 $\mu$m emission (free-free
emission).

\section{Prospects and expected developments}
Although optical interferometry has undergone significant improvements
in the past few years that have enabled to increase our understanding
of the inner part of protoplanetary disks on a growing sample of young
stars, some instrumental limitations yet prevent to unambiguously draw
a comprehensive picture of their environment. Strong efforts are 
now undertaken  to improve the capabilities of current and future
interferometers, which can
 be summarized around three main axes:\\
- {\it increasing the flux sensitivity:} thanks to dedicated fringe
  tracking/phase referencing devices, a better sensitivity will enable to
  observe (i) lower mass sources and (ii) fainter NIR emission lines
  which will be fully complementary of the $\mathrm{Br}\gamma$ line to
  characterize the rotating gas and the accretion/ejection phenomena
  (CO, Fe, Pa $\beta$,...).\\
- {\it going to higher spectral resolution:} a better sensitivity
  will also allow to use spectrographs performing higher resolution,
  with conserving enough flux in each spectral channel. Typically, a
  spectral resolution of $R > 8000$ (i.e. a few tens of km/s) will
  enable to spatially and {\it spectrally} resolve the emission lines,
  and will as well provide the velocity maps of their emitting regions, 
  putting to a test e.g. the rotation at the base of the jets
  \citep{bacciotti_1}, or the keplerian nature of the rotating gas. \\
- {\it developing the imaging capabilities:} an increased number
  of telescopes simultaneously  recombined (typically 6 or more
  \citep{filho_1}) is indeed mandatory to obtain enough measurements
  to obtain snapshot (i.e. in a few nights of observations at most)
  images of the inner environment of young stars, allowing to (i) set
  free from simple geometrical model-dependant analysis of the
  physical mechanisms at play and (ii) perform a temporal follow up of
  sources in
  adequacy with the dynamics at stake within the first AU of the protoplanetary disks.

\end{document}